\documentclass[%
 reprint,
 amsmath,amssymb,
 aps, prd
]{revtex4-2}
\usepackage{xcolor}
\usepackage{graphicx}
\usepackage{dcolumn}
\usepackage{bm}
\usepackage{hyperref}
\usepackage{multirow}
\begin{document}

\preprint{APS/123-QED}

\title{Engineering Confining Dilatons: A WKB Inverse Problem in Holographic QCD}

\author{Miguel Angel Martin Contreras}
\email{miguelangel.martin@usc.edu.cn}
\affiliation{
 School of Nuclear Science  and Technology\\
 University of South China\\
 Hengyang, China\\
 No. 28, West Changsheng Road, Hengyang City, Hunan Province, China.
}
\affiliation{Key Laboratory of Advanced Nuclear Energy Design and Safety,\\ Ministry of Education, Hengyang, 421001, China}

\author{Alfredo Vega}%
 \email{alfredo.vega@uv.cl}
\affiliation{%
 Instituto de F\'isica y Astronom\'ia, \\
 Universidad de Valpara\'iso,\\
 A. Gran Breta\~na 1111, Valpara\'iso, Chile
}
\begin{abstract}
This work presents a WKB-based inverse problem approach within the framework of holographic bottom-up QCD to engineer confining dilatons from hadronic mass spectra. Starting from a general parameterization of nonlinear radial Regge trajectories, $M_n^2=a(n+b)^\nu$, we apply the Rydberg-Klein-Rees (RKR) formula to derive the large-z behavior of the corresponding holographic confining potential. This potential is inversely related to the dilaton field profile, leading naturally to a non-quadratic dilaton  $\Phi(z)=(\kappa\,z)^{2-\alpha}$, where the parameters ($\kappa$, $\alpha$) are uniquely determined by the spectral parameters ($a$,$\nu$). We successfully test this method by fitting the spectra of heavy quarkonia ($c\bar{c}$ and $b\bar{b}$), achieving good agreement with experimental data. Furthermore, we extend this formalism to describe the spectroscopy of tetraquark states by superimposing an additional potential term, derived from the Bethe-Salpeter equation for diquarks, onto the standard mesonic confining potential. This work establishes a powerful and flexible bottom-up framework for deriving confinement directly from spectral data, applicable to both conventional and exotic hadrons.
\end{abstract}

\maketitle


\section{\label{sec:intro} Introduction}
The AdS/CFT correspondence\cite{Aharony:1999ti} has revolutionized our understanding of strongly coupled gauge theories, offering a powerful gravitational dual to phenomena such as confinement. Within this framework, bottom-up holographic QCD models \cite{Boschi-Filho:2002wdj, Polchinski:2002jw, Karch:2006pv} have emerged as a particularly effective phenomenological tool. These models map the problem of calculating hadronic properties in QCD to solving classical equations of motion for bulk fields in a deformed Anti-de Sitter (AdS) space, where conformal invariance is broken to generate a discrete mass spectrum. The central object in this construction is the dilaton field, $\Phi(z)$, whose profile dictates the form of the confining potential and ultimately the structure of the resulting Regge trajectories.

The most successful and widely adopted implementation, the softwall model \cite{Karch:2006pv}, employs a quadratic dilaton $\Phi(z)=\kappa^2\,z^2$ which yields linear radial Regge trajectories, $M_n^2\sim n$, a feature well-established for light, unflavored mesons. However, this simplicity proves to be a limitation. Experimental data for heavy quarkonia, such as charmonium and bottomonium, clearly display a nonlinear behavior \cite{Chen:2018bbr, MartinContreras:2020cyg}, where the mass squared of excited states does not increase linearly with the radial quantum number. This deviation is understood in traditional approaches, such as the Bethe-Salpeter formalism \cite{Chen:2018bbr, Chen:2023cws}, to be connected to the heavy quark mass, which challenges the universality of the quadratic dilaton and necessitates a more general formulation for a comprehensive holographic description of hadron spectroscopy.

This manuscript addresses this challenge by inverting the spectroscopic problem. Rather than positing a specific dilaton and deriving the spectrum, we employ the WKB approximation, specifically the Rydberg-Klein-Rees (RKR) formula \cite{MartinContreras:2025mpj}, to derive the large-z behavior of the confining potential—and hence the dilaton profile—directly from a parameterized hadronic mass spectrum, $M_n^2=a(n+b)^\nu$. We demonstrate that a nonlinear spectrum naturally implies a non-quadratic dilaton of the form $\Phi(z)=\left(\kappa\,z\right)^{2-\alpha}$, effectively engineering the holographic dual from experimental input. This approach not only provides a natural explanation for the deviation from linearity in heavy quarkonia but also offers a systematic pathway to extend the holographic description to more complex systems, such as multiquark states, whose spectra are less understood and cannot be captured by simple quadratic potentials.

This manuscript is structured as follows. In Section \ref{sec:framework}, we review the foundational framework of hadrons in holographic models, establishing the relationship between the bulk dilaton field, the confining potential, and the resulting hadronic mass spectrum. Section \ref{sec:wkb} presents the core of our work: a WKB-based inverse problem approach. We employ the Rydberg-Klein-Rees formula to derive the confining potential, and hence the dilaton profile $\Phi(z) = (\kappa z)^{2-\alpha}$, directly from a parameterized hadronic spectrum $M_n^2 = a(n+b)^\nu$. We then apply this method to heavy quarkonia ($c\bar{c}$ and $b\bar{b}$), demonstrating its efficacy in reproducing their nonlinear Regge trajectories. Finally, in Section~\ref{sec:multiquark}, we extend this formalism to the spectroscopy of tetraquark states by constructing a composite holographic potential that augments the mesonic potential with an additional term derived from the Bethe-Salpeter equation for diquarks. Our results establish a powerful, phenomenologically-driven framework for engineering confinement in holographic QCD.

\section{Hadrons in holographic models}\label{sec:framework}
The recipe for modeling hadrons in AdS/CFT models starts with the so-called \emph{holographic dictionary} \cite{Maldacena:1997re, Witten:1998qj, Aharony:1999ti}. Dual bulk fields to hadrons are characterized by the scaling dimension of the operators that create hadrons at the conformal boundary. This rule applies to bottom-up and top-down models \cite{Karch:2006pv, Erdmenger:2007cm}. 

In general, mesons are defined by operators with rank $r$, having a structure given by $\mathcal{O}_r=\bar{Q}\,\Gamma_p\,\hat{f}_q\left[D_\mu\right]\, Q$, where color and flavor indices are omitted for the sake of notation, $\Gamma_p$ are possible structures made of Dirac matrices with rank $p$ and $\hat{f}_q[D_\mu]$ are derivative structures that account for orbital excitations in mesons. For example, for mesons with $J>2$, these structures are written as \emph{symmetric traceless} tensors of rank $q$ made with covariant derivatives $D_\mu$. Notice that $r=p+q$ for this high-spin scenario. We can write the scaling dimension for these operators as 

\begin{equation}
    \text{dim}\,\mathcal{O}_r\equiv\Delta=\Delta_{Q}+L,
\end{equation}

\noindent where $L$ is the highest contribution of angular orbital momentum to the total spin of the meson $J=L+S$, where $S$ is the contribution of the quark-antiquark spin, and $\Delta_Q$ is the scaling dimension of quark clusters in the hadron. 

According to the field/operator duality, this operator of rank $r$ creating hadrons at the boundary is equivalent to an $r$-tensor bulk field living in the AdS$_5$ bulk. In general, for boson hadrons, we can write the most general action in terms of \emph{symmetric traceless} tensor field of rank $r$ as \cite{Gutsche:2011vb}

\begin{multline}\label{high_S}
    I_\text{Hadron}=-\frac{(-1)^r}{2\,\mathcal{K}}\,\int{d^{5}x\,\sqrt{-g}\,e^{-\Phi(z)}}\times\\
    \times\left[\nabla_M\,A^{N_1\ldots N_r}\,\nabla^M\,A_{N_1\ldots N_r}\right. \\
    \left.-M_{5}^2\,A^{N_1\ldots N_r}\,A_{N_1\ldots N_r}\right].
\end{multline}

\noindent where $\mathcal{K}$ is a coupling that fixes units in the bulk action and plays an important role in computing the decay constants, and it is fixed at the conformal boundary \cite{Erlich:2005qh}, $g_{mn}$ is the AdS$_5$ Poincarè patch metric, defined as

\begin{equation}
    dS^2=\frac{R^2}{z^2}\left[dz^2+\eta_{\mu\nu}\,dx^\mu\,dx^\nu\right]=\frac{R^2}{z^2}\,\eta_{MN}\,dx^M\,dx^N,
\end{equation}

\noindent with $R$ as the AdS radius. Regarding the index notation, the Greek indices $\mu,\,\nu,\ldots$ correspond to 4-dimensional quantities, and the capital Latin indices $M,\, N,\ldots$ correspond to 5-dimensional objects. 

The quantity $M_5$ is the bulk field mass responsible for defining the \emph{hadronic identity} encoding the dimension of the hadronic operator as

\begin{equation}
    M_5^2\,R^2=\left(\Delta-r\right)\left(\Delta+r-4\right),
\end{equation}

\noindent Thus, hadrons in holographic models are considered as \emph{bags of constituents}. There is no information about the inner constituent configuration \emph{ab initio}. 

The scalar bulk field $\Phi(z)$ is the dilaton, responsible for inducing \emph{confinement} in bottom-up models. The philosophy behind this approach is \emph{physical intuition}, motivated by the existing phenomenology that dictates how the dynamics within the bulk match the boundary physics. 

In the general bottom-up frame, the dilaton field is fixed \emph{phenomenologically}, considering the linearity of the Regge trajectories \cite{Karch:2006pv, MartinContreras:2020cyg}. However, we can extend the dilaton field to dynamical approaches where $\Phi(z)$  interacts with the background metric or other background fields to describe phenomena such as chiral symmetry breaking \cite{Batell:2008zm, Li:2013oda} or chemical and thermal effects relevant for phase dynamics \cite{Fang:2015ytf, Chen:2024mmd}. Nonetheless, the philosophy behind dynamic and static models is the same: the dilaton field induces confinement \emph{via} breaking the conformal invariance in the AdS bulk.

Let us pay attention to the equations of motion for the tensor field $A_{m_1 \ldots m_r}$. By varying the action, we obtain  

\begin{multline}
     \frac{1}{\sqrt{-g}}\nabla_M\left[\sqrt{-g}\,e^{-\Phi(z)}\,g^{MN}\,\nabla_N\,A^{M_1\ldots M_r}\right]\\
     -M_{5}^2\,e^{-\Phi(z)}\,A^{M_1\ldots M_r}=0,
\end{multline}

\noindent with $\nabla_M$ is the \emph{covariant derivative} written in terms of affine metric connections $\Gamma^M_{NP}$. However, it is customarily to assume the \emph{weak gravity limit}, in which $\Gamma^M_{NP}=0$ \cite{Gutsche:2011vb}. The gauge fixing in this scenario is the standard Coulomb-like one: 

\begin{equation}
    g^{M_1\,M_2}\,A_{M_1\,M_2\,\ldots M_r}=0,
\end{equation}

\noindent which implies

\begin{equation}
    \nabla^{M_1}\,A_{M_1 \dots M_r}=0.
\end{equation}

Therefore, after Fourier-transforming and introducing boundary sources as $A_{M_1\dots M_r}(z,q)=\tilde{A}_{M_1 \dots M_r}(q)\,\psi(z,q)$, we obtain the \emph{Sturm-Liouville form} of the equation of motion as

\begin{equation}
    \partial_z\left[e^{-B(z)}\,\partial_z\,\psi\right]+(-q^2)\,e^{-B(z)}\,\psi-\frac{M_5^2\,R^2}{z^2}\,e^{-B(z)}\,\psi=0,
\end{equation}

\noindent where we assume the on-shell mass condition $-q^2=M_n^2$ for hadronic masses, and we have defined the function $B(z)$ as

\begin{equation}
B(z)=\Phi(z)+\beta\,\log\left(\frac{R}{z}\right),
\end{equation}

\noindent that carries the dilaton and geometry information, as well as the rank of the bulk tensor field $r$ in the $\beta$ parameter: 

\begin{equation}
    \beta=-3+2\,r.
\end{equation}

We identify the hadronic spin in this step with $r=J$. In most bottom-up calculations, the $S-$ wave is assumed, \emph{ergo}, $J=S$ \cite{Erlich:2005qh, Karch:2006pv}. This $B(z)$ plays an important role in bottom-up models, as it defines the holographic confining potential. To do so, consider the Bogoliubov transformation   

\begin{equation*}
    u(z,q)=e^{\frac{1}{2}B(z)}\,\psi(z,q)
\end{equation*}

\noindent that transform the e.o.m in the Sturm-Liouville form into a \emph{Schrödinger-like} one:

\begin{equation}
    -u''+V(z)\,u=M_n^2\,u,
\end{equation}

\noindent with the confining potential $V(z)$ is written as

\begin{equation}
    V(z)=\frac{B'(z)^2}{4}-\frac{B''(z)}{2}+\frac{M_5^2\,R^2}{z^2}.
\end{equation}

Thus, the hadronic mass spectrum $M_n^2$ emerges as a radial Regge trajectory calculated as eigenvalues of the confining potential $V(z)$. Therefore, we can summarize the bottom-up program for hadron spectroscopy as: \emph{given a dilaton field (static or dynamic), define the holographic confining potential, and obtain the masses as eigenvalues of this potential.}

Let us comment on the confinement mechanism in the bottom-up approach. When there are no dilaton or geometric deformations, the AdS modes $\psi(z,q)$ are unbounded. In fact, the confining potential $V(z)$ is completely defined from the AdS warp factor, leading to a $1/z^2$ behavior. Two possibilities to avoid this behavior are to induce geometric deformations or include a dilaton. In the former, cutting the space with a D-brane results in an infinite square well potential, known as the hardwall model \cite{Polchinski:2002jw, Boschi-Filho:2002wdj, Erlich:2005qh}. Conversely, deforming the AdS space leads to confining terms in the confining potential \cite{FolcoCapossoli:2019imm}. In the latter case, the effect of the dilaton is the emergence of confining terms given by its derivatives: 

\begin{equation}\label{meson-pot}
    V(z)=\frac{\beta(\beta-2)+4\,M_{5}^2\,R^2}{4\,z^2}+V_\Phi\left[\Phi',\Phi''\right],
\end{equation}

\noindent where we have defined the pure dilaton part of the potential $V_\Phi\left[\Phi',\Phi''\right]$, which depends on the hadron spin information $\beta$, as

\begin{equation}\label{pot-phi}
    V_\Phi\left[\Phi',\Phi''\right]=-\frac{\beta}{2\,z}\Phi'(z)+\frac{1}{4}\Phi'(z)^2-\frac{1}{2}\Phi''(z).
\end{equation}

The high $z$ behavior of this potential, encoded in $V_\Phi\left[\Phi',\Phi''\right]$, induces the emergence of bounded states, identified with hadrons at the boundary. In the case of quadratic dilatons, \emph{a.k.a} the softwall model \cite{Karch:2006pv}, the resulting potential behaves as a harmonic potential, whose spectrum is radially linear $M_n^2=a(n+b)$, with $n$ the radial excitation number, expected for light unflavored hadrons. However, according to the Bethe-Salpeter theory, the concavity of Regge trajectories is connected to the quark mass. Therefore, for heavy hadrons, linearity ceases \cite{Chen:2018bbr}. A bottom-up approach to heavy quarkonia with nonlinear Regge trajectories was introduced in \cite{MartinContreras:2020cyg},  where a dilaton $\Phi(z)=(\kappa\,z)^{2-\alpha}$ was defined. In this framework, $\kappa$ is an energy scale that fixes the masses of the entire Regge trajectory, and $\alpha$ measures the deviation from the quadratic behavior. Thus, given the nonquadratic dilaton, the nonlinear Regge trajectory emerges. However, we will take the opposite approach, using WKB analysis in this manuscript, and conclude that for nonlinear Regge trajectories, the non-quadratic dilaton emerges naturally.  

\section{WKB approach in bottom-holography}\label{sec:wkb}
Motivated by the effect of dilaton derivatives on the hadron spectrum at the high $z$ limit, we can try to solve the inverse problem: Given a hadronic radial spectrum $M_n^2$, from experiment or non-holographic phenomenology, it is possible to obtain an expression for the confining part of the holographic bottom-up potential, i.e., dilaton derivatives. To do so, we will follow the WKB method applied in the so-called \emph{Rydberg-Klein-Ross formula} \cite{Quigg:1979vr, MartinContreras:2025mpj}. 

Let us consider the following structure for nonlinear Regge trajectories, motivated by heavy-quarkonia phenomenology:

\begin{equation}\label{test-trajectory}
    M_n^2=a\left(n+b\right)^\nu,
\end{equation}

\noindent where $a$ is the Regge slope, $b$ the intercept and $\nu$ defines the divination from linearity. We learned from the original softwall \cite{Karch:2006pv} that the large $z$ behavior observed in the holographic potential naturally controls confinement. In the case of top-down approaches, the equivalent of large $z$ behavior is to \emph{conveniently} deform the bulk space and bulk fields to induce a mass tower of excited states dual to mesons, as in the D3/D7 approach \cite{Kruczenski:2003be}.  

The RKR formula \cite{Rees, Rydberg1, Rydberg2, Klein, RKRref} allows us to extract the behavior of the potential near a given turning point $z^*$ in terms of the spectrum $M_n^2$ and its derivative as follows:

\begin{equation}\label{RKR-for}
z(V^*)=2\int_0^{V^*}{\frac{d\,M^2}{\frac{d\,M^2}{d\,n}\left(V^*-M_n^2\right)^{1/2}}}.     
\end{equation}

Since the conformal boundary is at $z\to0$, for holographic bottom-up potentials, the turning points are $z\to0$ and $z(V^*)$, with the property that $z(V^*)\to\infty$, with $V^*$ defined as the potential near the turning point, \emph{i.e.,} $V(z^*)\equiv V*$. Thus, for the nonlinear Regge trajectory, we have

\begin{eqnarray}
z(V^*)&=&2\int_0^{V^*}{\frac{d\,M^2_n}{a\,\nu\,\left(\frac{M_n^2}{a}\right)^{\frac{\nu-1}{\nu}}\left(V^*-M_n^2\right)^{1/2}}}\\
&=&\frac{2\,\pi^{1/2}}{a^{1/\nu}}\frac{{V^*}^{\frac{2-\nu}{\nu}}\,\Gamma\left(\frac{\nu+2}{2\,\nu}\right)}{\nu\,\Gamma\left(\frac{1}{\nu}\right)}. 
\end{eqnarray}

Inverting this expression, we can find the high-$z$ potential as 

\begin{equation}
 V^*(z)=\left[\frac{a^{1/\nu}}{2\,\pi^{1/2}}\,\frac{\nu\,\Gamma\left(\frac{\nu+2}{2\,\nu}\right)}{\Gamma\left(\frac{1}{\nu}\right)}\,z\right]^{\frac{2\,\nu}{2-\nu}}\equiv C(\nu,a)\,z^{\frac{2\,\nu}{2-\nu}},
\end{equation}

\noindent where $0<\nu<2$, and 

\begin{equation}\label{const}
    C(\nu,\,a)=\left[\frac{a^{1/\nu}}{2\,\pi^{1/2}}\frac{\nu\,\Gamma\left(\frac{\nu+2}{2\,\nu}\right)}{\Gamma\left(\frac{1}{\nu}\right)}\right]^{\frac{2\,\nu}{2-\nu}}.
\end{equation}

We use $V^*$ to denote the potential calculated from the WKB from the holographic potential $V(z)$. Now, we can solve the dilaton for the nonlinear trajectory case as follows: 

\begin{equation}
C(\nu,a)\,z^{\frac{2\,\nu}{2-\nu}}=  \frac{1}{4}\Phi'(z)^2-\frac{1}{2}\Phi''(z)-\frac{\beta}{2\,z}\Phi'(z)
\end{equation}

\noindent where we fix the boundary conditions as:

\begin{eqnarray}
    \Phi'(z\to0)&=&0,\\
    \Phi'(z\to \infty)&=&\frac{2\,\nu}{2-\nu}\,C(\nu,a)\,z_\infty^{\frac{2\,\nu}{2-\nu}-1},    
\end{eqnarray}

\noindent where $z_\infty$ is a large $z$ value used to fit the numerical infinity. We use this recipe to infer the dilaton associated with the Regge trajectory \eqref{test-trajectory}. Thus, we write the holographic potential as: 

\begin{equation}
    V(z)=\frac{\beta(\beta-2)+4\,M_{5}^2\,R^2}{4\,z^2}+V^*(z).
\end{equation}

As a hypothesis, let us define the static dilaton profile as $\Phi(z)=\left(\kappa\,z\right)^\gamma$, where $\gamma=2$ recovers the standard softwall model. The dominant terms in the holographic potential at large $z$ come from the $\Phi'(z)^2$ term in the potential. Thus, we can approximately meet the following condition:  

\begin{equation}\label{WKB_cond}
    V^*(z)\approx\frac{1}{4}\Phi'(z)^2\Rightarrow\frac{1}{4}\, \gamma^2\,\kappa^{2\gamma}\,z^{2(\gamma-1)}= C(\nu,a)\, z^{\frac{2\,\nu}{2-\nu}}.
\end{equation}

By requiring that both parts match, we will obtain the following relations for the exponent $\gamma$ and the energy scale $\kappa$ as:

\begin{eqnarray}
\gamma&=&\frac{2}{2-\nu}\\
\kappa &=&\left[\left(2-\nu\right)^2\,C(\nu,a)\right]^{\frac{2-\nu}{4}},
\end{eqnarray}

\noindent where $C(\nu,\,a)$ was defined in the expression \eqref{const}. Notice that fixing $\nu=1$ will recover the softwall model again:

\begin{eqnarray}
    \gamma&=&2\\
    \kappa=\left[C\left(1,a\right)\right]^{1/4}&=&\left[\frac{a}{2}\right]^{1/2},
\end{eqnarray}

\noindent thus we have for the dilaton $\Phi(z)=\kappa^2\,z^2,$ and for the radial trajectory $M_n^2=4\,a(n+b)$, where $b$ should be determined from the low-$z$ behavior in the potential controlled by the bulk mass $M_5$ and the hadron spin $J$ \cite{Afonin:2017zis}. 

Let us test this model with heavy vector quarkonia. To do so, let us substitute $\gamma=2-\alpha$, where $\alpha$ measures the deviation from the softwall model case. Thus, nonlinearities in the radial spectrum will be translated into deviations from the quadratic behavior of the static dilaton. Therefore, 

\begin{eqnarray} \label{kappa-nonlin}
  \kappa &=&\left[\frac{a^{1/\nu}}{2\,\pi^{1/2}}\frac{\nu\,\Gamma\left(\frac{\nu+2}{2\,\nu}\right)}{\Gamma\left(\frac{1}{\nu}\right)}\right]^{\frac{\nu}{2}}\left(2-\nu\right)^{\frac{2-\nu}{2}}\\ \label{alpha-nonlin}
  \alpha&=&\frac{2\left(1-\nu\right)}{2-\nu}. 
\end{eqnarray}

Let us use these expressions in the heavy vector meson spectra. We chose these states because they provide nonlinear fits for their experimental radial Regge trajectories, as summarized in Table \ref{tab:one}.

\subsection{Heavy Quarkonia}

To test the efficiency of the WKB approach, we will consider the heavy quarkonium vector spectra, given by the $c\bar{c}$ and $b\bar{b}$ states. For these states, the spectroscopy systematics is written as $n^{2S+1}
\,L_J$, with $J=L+S$, $P=(-1)^{L+1}$, $C=(-1)^{L+S}$, $G=(-1)^{L+S+I}$. From a Bethe-Salpeter analysis, these heavy mesons are described by nonlinear trajectories \cite{Chen:2018bbr, Chen:2018hnx, Chen:2018nnr}. The deviation from the linearity observed in light unflavored mesons is related to the quark mass. From an NRQCD analysis \cite{Lucha:1991vn, Mony2023bxv},  it is possible to derive a relation connecting the hadron wave function at the origin $\Psi(0)$ with the masses of the spin-singlet and spin-triplet meson states as

\begin{equation}
   \Delta M^2= M^2_{S_1}-M^2_{S_0}=\frac{32\,\pi\,\alpha_s}{9\,m_q}\left|\Psi(0)\right|^2.
\end{equation}

Assuming an NRQCD potential such as Cornell, it is possible to infer that, for light mesons, $\Delta M^2$ does not vary, implying that the confining linear term is dominant. However, it is the opposite for heavy mesons: in these systems, the heavy quarks are more influenced by the Coulomb-like part of the potential. Thus, heavy ground states should be significantly more affected by the Coulomb-like term than excited ones. This fact is reflected in the high difference between the ground state and the first excited state masses compared to the difference between the masses of any two excited states, giving rise to the nonlinearity in the Regge trajectories.  

We summarize the meson masses for the ground and excited states, as well as the nonlinear Regge trajectory parametrization coefficients, in Table \ref{tab:one}.  With these data, we will compute, using the WKB formulas for $\kappa$ and $\alpha$, e.g., expressions \eqref{kappa-nonlin} and \eqref{alpha-nonlin}, the dilaton fields for charmonium and bottomonium, as follows:

\begin{eqnarray}\label{cc-par}
   \Phi_{c\bar{c}}(z)&=&\left[\left(2.174\pm0.585\right)\,z\right]^{2-\left(0.561\pm0.176\right)} \\ \label{bb-par}
   \Phi_{b\bar{b}}(z)&=&\left[(12.832\pm0.723)\,z\right]^{2-(0.893\pm0.018)},
\end{eqnarray}

\noindent with all $\kappa$ scales given in GeV. 

\begin{center}
\begin{table*}[t]
    \begin{tabular}{|c||c|c|c|c|c|c|}
 \hline
    \multicolumn{7}{||c||}{\textbf{Heavy Quarkonia summary in MeV $I^G(J^{PC})=0^+(1^{--})$}}\\
    \hline
    \hline
    \textbf{$Q\bar{Q}$ state} &  $1^3\,S_1$ & $2^3\,S_1$ & $3^3\,S_1$ & $4^3\,S_1$ & $5^3\,S_1$& $6^3\,S_1$ \\
    \hline
    \multirow{2}{4em}{\centering\textbf{$c\bar{c}$}} & $3096.900\pm0.006$, & $3686.097\pm0.011$, & $4040\pm4$,& $4415\pm5$ &  Not Seen & Not Seen \\
    \cline{2-2} \cline{3-3} \cline{4-4} \cline{5-5}\cline{6-6}\cline{7-7}
    & \multicolumn{2}{c|}{Non-Linear Regge Trajectory:} & \multicolumn{1}{c}{$a(\text{GeV}^2)=7.88\pm 2.62$}, & \multicolumn{1}{c}{$b=0.39\pm0.67$}, & $\nu=0.61\pm0.17$, & $\mathcal{R}^2=0.999$\\
    \hline
    \hline
    \multirow{2}{4em}{\centering\textbf{$b\bar{b}$}} & $9460.40\pm0.10$ & $10023.4\pm0.5$ & $10355.1\pm0.5$ & $10579.4\pm1.2$ & $10885.2^{+2.6}_{-1.6}$  & $11000\pm4$\\
    \cline{2-2} \cline{3-3} \cline{4-4} \cline{5-5}\cline{6-6}\cline{7-7}
    & \multicolumn{2}{c|}{Non-Linear Regge Trajectory:} & \multicolumn{1}{c}{$a(\text{GeV}^2)=84.98\pm 5.34$}, & \multicolumn{1}{c}{$b=0.32\pm0.41$}, & $\nu=0.193\pm0.029$ &  $\mathcal{R}^2=0.999$\\
    \hline
    \hline
    \end{tabular}
    \caption{This table summarizes the experimental masses of heavy vector quarkonia and their systematic spectroscopic notation and quantum numbers \cite{ParticleDataGroup:2024cfk}. We also present the nonlinear Regge trajectory parametrization, given in equation \eqref{test-trajectory}, along with the correlation coefficient $\mathcal{R}^2$, which is necessary to compute the values of the parameters $\kappa$ and $\alpha$ for each heavy quarkonium spectrum.}
    \label{tab:one}
\end{table*}    
\end{center} 

Regarding the radial spectra used, we identify as radial excited states of the $c\bar{c}$ system the $\psi(4040)$ as $3^3\, S_1$ and $\psi(4415)$ as $4^3\, S_1$ because the remaining states have properties consistent with exotic hadrons. In a similar scenario for $b\bar{b}$ states, we identify $\Upsilon(10860)$ as $5^3\,S_1$ and $\Upsilon(11020)$ as $6^3\,S_1$ \cite{Cleven:2015era, Husken:2024rdk}.

We can compute the holographic spectra with static dilatons \eqref{cc-par} and \eqref{bb-par} for the radial and angular states observed in $c\bar{c}$, and $b\bar{b}$. Recall that the holographic confining potential is written as

\begin{eqnarray}\label{qqbar-pot}
    V_{Q\bar{Q}}(z)&=&\frac{\beta(\beta-2)+4\,M_{5}^2\,R^2}{4\,z^2}+V_\Phi\left[\Phi',\Phi''\right]\\
    M_5^2\,R^2&=&\left(3+L-J\right)\left(L+J-1\right)\\
    \beta&=&-3+2\,J,
\end{eqnarray}

\noindent where the pure dilaton part of the potential $V_\Phi$ is defined in the expression \eqref{pot-phi}. Tables \ref{tab:two}
and \ref{tab:three} summarizes the holographic results for the charmonium and bottomonium spectra.

The higher differences (relative error) with the experimental values for charmonium spectra occur in the pseudoscalar ground state $\eta_c(1S)$ and for the lowest orbital states in the $P$ and $D$ trajectories. From a holographic point of view, such differences arise because the low $z$ limit of the confining potential does not adequately capture NRQCD effects \cite{Braga:2015jca, Braga:2017bml}. 

To quantify the accuracy of this model, we consider the Root-Mean-Square error (RMS) for a set of experimental data with $N$ entries, $N_p$ number of parameters, defined as:

\begin{equation}\label{RMS}
    \Delta_\text{RMS}=\sqrt{\frac{1}{N-N
    _p}\,\sum_{n=1}^{N}\left(\frac{M_{n,\,\text{Exp}}-M_{n,\,\text{Theo}}}{M_{n,\,\text{Exp}}}\right)^2}\times\,100.
\end{equation}

\noindent that measures the deviation observed in the theoretical outcomes from the experimental data. 

Despite this fact, this holographic model, with two parameters $\kappa_c$ and $\alpha_c$,  has an RMS error of $12.2\%$ when fitting 17 hadronic masses. 

For the bottomonium case, a similar story occurs. The highest differences emerge for the pseudoscalar ground state $\eta_b(1S)$ and the lowest states in the orbital trajectories $P$ and $D$. In this particular situation, eqn. \eqref{RMS} yiedls, for fitting 19 masses with two parameters, $\kappa_b$ and $\alpha_b$, a $7.7\%$ RMS error. 

\begin{center}
\begin{table*}[t]
    \begin{tabular}{|c|c|c|c|c|c|}
    \hline
    \hline
    \multicolumn{6}{c}{\textbf{Summary of Charmonium states}}\\
    \multicolumn{3}{c}{$\kappa_c=2.174$ GeV}&\multicolumn{3}{c}{$\alpha_c=0.561$}\\
    \hline
    \hline
    \textbf{State} &\textbf{$J^{PC}$}&\textbf{$n^{2S+1}\,L_J$}&\textbf{$M_\text{Exp}$ (MeV)}&\textbf{$M_\text{Th}$ (MeV)}& $\Delta\,M(\%)$\\
    \hline
    \hline
    \multirow{2}{2em}{$\eta_c$}&\multirow{2}{*}{$0^{-+}$}&$1^1\,S_0$&$2984\pm0.4$&$3633.9$&$21.8$\\
    & & $2^1\,S_0$&$3637.7\pm0.9$&$4051.8$&$11.4$\\
    \hline
    \multirow{8}{*}{$\psi$}&\multirow{6}{*}{$1^{--}$}&$1^3\,S_1$&$3096.097\pm0.006$&$3042.75$&$1.74$\\  
    & & $2^3\,S_1$& $3686.097\pm0.011$ & $3619.8$ & $1.80$\\
    & & $3^3\,S_1$& $4040\pm4$ & $4041.0$ & $0.1$ \\
    & & $4^3\,S_1$& $4415\pm5$ & $4380.7$ & $0.8$\\ \cline{3-3} \cline{4-4} \cline{5-5}\cline{6-6}
    & &$1^3\,D_1$ & $3773.7\pm0.7$ & $3909.9$ & $3.61$\\
    & & $2^3\,D_1$ & $4191\pm5$ & $4264.6$ &$1.8$\\
    \cline{2-2} \cline{3-3} \cline{4-4} \cline{5-5}\cline{6-6}
    & $2^{--}$ & $1^3\,D_2$ & $3823.51\pm0.34$ & $3212.7$ &$15.98$\\ \cline{2-2} \cline{3-3} \cline{4-4} \cline{5-5}\cline{6-6}
    & $3^{--}$ & $1^3\,D_3$& $3842.71\pm0.20$ & $3282.7$  &$14.57$\\
    \hline
    \multirow{6}{*}{$\chi_c$} & \multirow{3}{*}{$0^{++}$}& $1^3\,P_0$ & $3414.71\pm0.30$ & $3852.5$&$12.82$\\
    & & $2^3\,P_0$& $3862^{+50}_{-35}$ & $4228.9$ & $9.51$\\ 
    & & $3^3\,P_0$& $3922.1\pm1.8$ & $4508.8$ & $15.77$\\\cline{2-2} \cline{3-3} \cline{4-4} \cline{5-5}\cline{6-6}
    & $1^{++}$ & $1^3\,P_1$ &$3510.67\pm0.05$ & $3387.3$ & $3.51$\\ \cline{2-2} \cline{3-3} \cline{4-4} \cline{5-5}\cline{6-6}
    &\multirow{2}{*}{$2^{++}$}& $1^3\,P_2$ & $3556.17\pm0.07$ & $2807.8$ & $21.04$\\
    & & $2^3\,P_2$ & $3922.5\pm1.0$ &$3430.9$&$12.53$\\ \cline{2-2} \hline
    $h_c$ &$1^{+-}$ &$1^1\,P_1$ & $3525.37\pm0.14$ & $3387.3$ & $3.91$\\
    \hline
    \hline
    \multicolumn{6}{c}{RMS error (17 states, 2 parameters)$=12.18\%.$}
    \end{tabular}
    \caption{Summary of the charmonium spectroscopy for $q\bar{q}$ states computed as eigenvalues of the holographic confining potential \eqref{qqbar-pot}. Experimental data is read from 
 PDG \cite{ParticleDataGroup:2024cfk}. }
    \label{tab:two}
\end{table*}    
\end{center} 

\begin{center}
\begin{table*}[t]
    \begin{tabular}{|c|c|c|c|c|c|}
    \hline
    \hline
    \multicolumn{6}{c}{\textbf{Summary of Bottomonium states}}\\
    \multicolumn{3}{c}{$\kappa_b=12.832$ GeV} &\multicolumn{3}{c}{$\alpha_b=0.893$}\\
    \hline
    \hline
    \textbf{State} &\textbf{$J^{PC}$}&\textbf{$n^{2S+1}\,L_J$}&\textbf{$M_\text{Exp}$ (MeV)}&\textbf{$M_\text{Th}$ (MeV)}& $\Delta\,M(\%)$\\
    \hline
    \hline
    \multirow{2}{*}{$\eta_b$}&\multirow{2}{*}{$0^{-+}$}&$1^1\,S_0$&$9398.7\pm2.0$&$10785.1$&$14.7$\\
    & & $2^1\,S_0$&$9999\pm4$&$10981.8$&$9.82$\\
    \hline
    \multirow{7}{*}{$\Upsilon$}&\multirow{7}{*}{$1^{--}$}&$1^3\,S_1$&$9460\pm0.10$&$9942.9$&$5.10$\\  
    & & $2^3\,S_1$& $10023.4\pm0.5$ & $10320.5$ & $2.96$\\
    & & $3^3\,S_1$& $10355.1\pm0.5$ & $10598.7$ & $2.35$ \\
    & & $4^3\,S_1$& $10579.4\pm1.2$ & $10820.8$ & $2.28$\\
    & & $5^3\,S_1$& $10885^{2.6}_{1.6}$ & $11006.4$ & $1.11$\\
    & & $6^3\,S_1$& $11000\pm4$ & $11166.3$ & $1.51$\\ 
    \cline{2-2} \cline{3-3} \cline{4-4} \cline{5-5}\cline{6-6}
    & $2^{--}$ & $1^3\,D_2$ & $10163.7\pm1.4$ & $9496.8$ &$6.56$\\ 
    \hline
    \multirow{6}{*}{$\chi_b$} & \multirow{2}{*}{$0^{++}$}& $1^3\,P_0$ & $9859.44\pm0.42\pm0.31$ & $10849.1$&$10.04$\\
    & & $2^3\,P_0$& $10232.5\pm0.4\pm0.5$ & $11036.1$ & $7.85$\\ \cline{2-2} \cline{3-3} \cline{4-4} \cline{5-5}\cline{6-6}
    & \multirow{3}{*}{$1^{++}$} & $1^3\,P_1$ &$9892.78\pm0.26\pm0.31$ & $10900.3$ & $10.18$\\ 
    & & $2^3\,P_1$ &$10255.46\pm0.22\pm0.50$ & $11079.6$ & $8.04$\\
    & & $3^3\,P_1$ &$10513.4\pm0.7$ & $11234.2$ & $6.86$ \\
    \cline{2-2} \cline{3-3} \cline{4-4} \cline{5-5}\cline{6-6}
    &\multirow{3}{*}{$2^{++}$}& $1^3\,P_2$ & $9912.21\pm0.26\pm0.31$ & $10915.6$ & $10.12$\\
    & & $2^3\,P_2$ & $10268.65\pm0.22\pm0.50$ &$11092.6$&$8.03$\\ 
    & & $3^3\,P_2$ & $10524.0\pm0.8$ &$11245.5$&$6.86$\\ \cline{2-2} \hline
    \multirow{2}{*}{$h_b$} &\multirow{2}{*}{$1^{+-}$} &$1^1\,P_1$ & $9899.3\pm0.8$ & $10155.9$ & $2.59$\\
    & &$2^1\,P_1$ & $10259.8\pm1.2$ & $10471.2$ & $2.06$\\
    \hline
    \hline
    \multicolumn{6}{c}{RMS error (19 states, 2 parameters)$=7.67\%.$}
    \end{tabular}
    \caption{Summary of the bottomonium spectroscopy for $q\bar{q}$ states computed as eigenvalues of the holographic confining potential \eqref{qqbar-pot}. Experimental data is read from 
 PDG \cite{ParticleDataGroup:2024cfk}.}
    \label{tab:three}
\end{table*}    
\end{center} 

\section{Multiquark states using bottom-up}\label{sec:multiquark}
Let us consider the extension of these WKB ideas to describe multiquark states. In the context of holographic QCD, the spectroscopy of such states has been addressed in Sakai-Sugimoto, deformed backgrounds, and dilaton-like models \cite{Forkel:2010gu, Liu:2019yye, Dosch:2020hqm, MartinContreras:2023oqs, Toniato:2025gts, Karapetyan:2021ufz}.

We will propose an alternative approach based on WKB analysis by considering a possible form for the Regge trajectories expected for tetraquark structures. We will restrict ourselves to the case of heavy vector exotic states.

Recall that the so-called \emph{constituent quark model} describes the observed spectrum of $q\bar{q}$ bounded states, mapping them into $SU(N)$ flavor multiplets. This description enables us to establish a taxonomy for hadrons, comprising conventional mesons, conventional baryons, gluonic excitations, and multiquark states (including compact, hadroquarkonium, and molecular states). The latter two categories do not fit this schematic description. These exotic states are classified according to whether quantum numbers are allowed or not in the context of the quark constituent model. 

From a dynamical perspective, strong residual forces arise from the interaction between non-conventional constituent clusters (e.g., quarks, gluons, meson cores). However, the difference resides \emph{in the energy scale} in each scenario: For \emph{compact multiquark} states, multiquark clusters interact with gluon exchange at short range. For \emph{hadroquarkonium}, a Van der Waals-color-like interaction is expected to polarize the surrounding meson cloud due to the presence of a heavy meson core. And for \emph{hadronic molecules}, the interaction is due to meson-exchange, implying a long range.  

However, this approach also implies revising how we introduce multiquark operators in the holographic dictionary. Recall that from the field/operator duality, beyond the twist, there is no \emph{ab initio} information about the hadron inner structure or compactness. For compact multiquark structures, we expect dimension-six quark operators \cite{Dudek:2010wm} with local effects encoded into contracted differential operators \cite{Cheung:2017tnt}. 

On the holographic side, including covariant contracted derivatives (e.g., Laplacians) in the multiquark operators adds local information related to the inner structure, without changing the hadronic operator rank. Therefore, we will consider the tetraquark composed as a pair of diquark-antidiquark interacting, $[Q_1\,Q_2\,][\bar{Q_3}\,\bar{Q_4}]$. These structures will be associated with $r-$ rank boundary hadronic operators defined as \cite{Afonin:2020msa}

\begin{equation}
 \mathcal{O}^{4Q}_r=\bar{Q}_1\,\Gamma_p\,D^{2i}\,Q_2\,\bar{Q}_3\,\Gamma_q\,D^{2i}\,Q_4   
\end{equation}

\noindent where $\Gamma_{p,q}$ are structures written with 4-dimensional Dirac gamma matrices, $D^2=\eta_{\mu\nu}\,D^\mu\,D^\nu$ is the Laplacian defined with gauge boundary covariant derivatives, $Q_i$ are the quark fields. The integers $\{i,k\}\in \mathbb{Z}^+\cup\{0\}$ count for fully contracted derivatives that do not change rank $r$ but twist, defined as 

\begin{equation}
    \text{dim}\,\mathcal{O}_r=\Delta_q+2\left(i+j\right)-r.
\end{equation}


The redefinition of the scaling dimension implies that $M_5^2\, R^2$ changes its value; it is no longer zero, as we are no longer considering regular vector mesons. 

These changes do not alter the functional structure observed for the confining potential, as given by the equation. \eqref{qqbar-pot}. Therefore, $V_{Q\,\bar{Q}}(z)$ captures partially the multiquark confinement scenario. In terms of the dilaton field, which accounts for the strong interaction that binds the quarks inside the meson, the set of parameters $\{\kappa,\,\alpha\}$ is not enough to account for the \emph{residual strong force} that binds multiquark structures. A natural form to include such effects is by adding an extra dilaton field $\tilde{\Phi}(z)$. Thus, we define

\begin{equation}
    \Phi_{4Q}(z)=\left(\kappa\,z\right)^{2-\alpha}+\tilde{\Phi}(z).
\end{equation}

This dilaton superposition defines a confining potential with two main parts. One coming from the non-quadratic dilaton $\left(\kappa\,z\right)^{2-\alpha}$ having the same structure as $V_{Q\bar{Q}}(z)$ but a different $M_5^2\,R^2$. The second part, which we will call $\tilde{V}(z)$, is defined by the derivatives of $\tilde{\Phi}(z)$ and encodes the multiquark effect. 

How will $\tilde{V}(z)$ be defined? We will use the tetraquark phenomenology and the RKR formula to construct it. Our source will be the Bethe-Salpeter theory for multiquark states. In App. \ref{app2}, we deliver a brief introduction to the main ideas behind the quadratic spinless Bethe-Salpeter equation (QSSE) for heavy diquarks. 


In the work \cite{Feng:2023txx}, the authors explore the use of the \emph{quadratic form of the spinless Bethe-Salpeter equation} \cite{Baldicchi:2007ic} to explore the Regge trajectories associated with diquark structures. Recall that tetraquark states can be considered as a pair of diquark-antidiquark. Fully heavy diquarks, i.e. $[Q_1,Q_2]$ or $[\bar{Q_1},\bar{Q_2}]$, can be described as two color representations under $SU(3)_c$: the \emph{attractive} channel given the antitriplet $\bar{\textbf{3}}_c$, and the \emph{repulsive} corresponding with the sextet $\textbf{6}_c$. In forming \emph{hiddend-charm tetraquarks}, we will consider the former representation. This representation also holds when we consider heavy-light diquarks, as in the case of \emph{open-charm tetraquarks.}

In the heavy case, the Regge trajectories for the diquarks have the following asymptotic form for radial excitations \cite{Feng:2023txx}:

\begin{equation}
    M_n^2\sim (3\,\pi)^{2/3}\left(m_1+m_2\right)\left(\frac{\sigma_c^2}{4\,\mu}\right)^{1/3}\,n^{2/3},
\end{equation}

\noindent where $m_1$ and $m_2$ are the quark constituent masses in the diquark structure, $\mu$ is the reduced mass in the diquark, and $\sigma_c$ is the string tension in the Cornell potential for charmonia \cite{Lucha:1991vn}. This Regge trajectory allows us to use the RKR formula \eqref{RKR-for} to extract the potential, obtaining:

\begin{equation}
    \tilde{V}(z)=2\,g_\text{eff}\,\left(m_1+m_2\right)^{3/2}\left(\frac{\sigma_c^2}{4\,\mu}\right)^{1/2}z
\end{equation}

\noindent where $g_\text{eff}=\frac{1}{2\pi}$ is constant accounting for the low $n$ effects, and it is fixed numerically by minimizing the RMS error. In appendix \ref{ap1}, we explain how to compute the $\sigma_c$ string tension from the Cornell potential.  

Therefore, we write  the full holographic confining potential for tetraquarks as

\begin{equation}
    V_{4Q}(z)=V_{q\bar{q}}\left[z,\tilde{M}_5^2\,R^2\right]+\tilde{V}(z),
\end{equation}

\noindent where the bulk mass $\tilde{M}_5^2\,R^2$ accounts for the total quark content, the inner structure, and the hadronic spin:

\begin{equation}
    \tilde{M}_5^2\,R^2=\left[\Delta_{4Q}+2\left(i+j\right)-J\right]\left[\Delta_{4Q}+2\left(i+j\right)-J-4\right].
\end{equation}

Notice that the $\tilde{M}_5^2\,R^2$ structure implies that the tetraquark states are described as $S$ states. The $i,j$ indices account for non-local effects due to color dynamics. It is possible to include angular momentum effects, associated with molecular states, as loosely bound $D\,\bar{D}^*$ mesons. However, we will follow the hypothesis that these tetraquark structures, composed of diquark-antidiquark cores, are compact in space; therefore, it is expected to find them in $S$ states \cite{Sazdjian:2022kaf}. This tetraquark picture is also consistent with the holographic interpretation of hadrons as punctual bags of constituents, characterized by the dimension of the operator creating such hadrons. 

Regarding the field/operator duality, notice that for four constituent quarks, the scaling dimension is fixed to be $\Delta_{4Q}=6$. In general, this parameter does not distinguish among exotic structures, such as molecular or 
compact tetraquarks. 

To perform numeric analysis, we need to set four parameters: $\kappa$ and $\alpha$ for the $V_{q\bar{q}}(z)$, and the set $g_{eff}$ and $\sigma_c$ for the WKB potential. The latter is fixed from the Cornell potential. $g_{eff}$ is chosen in such a way that it adjusts the low $n$ behavior for Regge trajectories. Since we have four quarks, organized in a diquark-antidiquark structure, we can analyze the cases for \emph{open-charm} and \emph{hidden-charm} tetraquarks, i.e., $[Q_1\,q_2][\bar{Q}_3\,\bar{q}_4]$ and $[Q_1\,Q_2][\bar{q}_3\,\bar{q}_4]$. This clusterization also serves as constraints for fixing the remaining parameters: one of the clusters will feed $V_{q\bar{q}}(z)$, allowing to fix $\kappa$ and $\alpha$, from the stoichiometric running of these parameters with the constituent quark mass (see Appendix \ref{Appx2} for further details). 

While this extension introduces additional parameters ($g_{eff}$, $\sigma_c$, $i$, $j$) beyond the mesonic ($\kappa$, $\alpha$) set, they are not free in the usual sense. The string tension $\sigma_c$ is determined from the Cornell potential (Appendix A), and $g_{eff}$ is fixed to a constant value $\frac{1}{2\pi}$ derived from the WKB analysis of the diquark Bethe-Salpeter equation. The indices $i$ and $j$, which account for internal Laplacian operators, are discrete integers chosen based on the exotic state quantum numbers. This structured approach mitigates the risk of overfitting despite the increased model complexity.

\begin{center}
\begin{table*}[t]
    \begin{tabular}{|c|c|c|c|c|c|c|c|}
    \hline
    \hline
    \multicolumn{6}{c}{\textbf{Summary of Charmonium Tetraquarks}}\\
    \multicolumn{3}{c}{$\sigma_c=0.24$ GeV$^2$}&\multicolumn{3}{c}{$g_{eff}=
    \frac{1}{2\pi}$}\\
    \hline
    \hline
    \textbf{State} &\textbf{$J^P$ or $J^{PC}$}&\textbf{$[q_1\,q_2][\bar{q}_3\,\bar{q}_4]$}&\textbf{$M_\text{Exp}$ (MeV)}&\textbf{$M_\text{Th}$ (MeV)}& $\Delta\,M(\%)$ & $(i,j)$\\
    \hline
    \hline
    $T^*_{cs0}(2870)^0$ & $0^+$ &$[c d][\bar{s}\bar{u}]$ &$2866\pm7$& $2704.9$ & $5.62$ &$(0,0)$\\
    \hline
    $T^*_{cs1}(2900)$& $1^-$ & $[su][\bar{c}\bar{d}]$&$2904\pm4$&$2943.5$ & $1.4$&$(0,0)$\\
    \hline
    \multirow{2}{*}{$T_{cc}(3875)^+$}&$1^+$&\multirow{2}{*}{$[c c][\bar{u}\bar{d }]$}& $3874.83\pm0.11$&$3990.32$ & $2.98$ & $(1,1)$\\
    &$0^+$& & $3874.83\pm0.11$ &$3846.2$& $0.74$ & $(0,1)$\\
    \hline
    $T_{c\bar{c}1}(3900)$&$1^{+-}$&$[cu][\bar{c}\bar{u}]$&$3887.1\pm2.6$ &$3918.3$ & $0.80$ &$(0,0)$\\
    \hline
    $T_{c\bar{c}\bar{s}1}(4000)$&\multirow{2}{*}{$1^{+}$}&\multirow{2}{*}{$[c u][\bar{c}\bar{s}]$}&$3980$ to $4010$ &$4133.9$ & $3.47$ &$\multirow{2}{*}{(0,0)}$\\
    $T_{c\bar{c}\bar{s}1}(4220)$& & &$4220^{+50}_{-40}$ &$4451.8$ & $5.49$ &\\
    \hline
    \multirow{2}{*}{$T_{c\bar{c}}(4050)^+$}&$1^{?+}$&\multirow{2}{*}{$[cu][\bar{c}\bar{d}]$}&\multirow{2}{*}{$4051^{+24}_{-40}$}&$4215.4$& $4.06$ & $(0,0)$\\
    &$0^{?+}$& & &$4331.8$& $6.94$& $(0,0)$\\
    \hline
    \multirow{2}{*}{$T_{c\bar{c}}(4055)^+$}&$1^{?-}$&\multirow{2}{*}{$[cu][\bar{c}\bar{d}]$}& \multirow{2}{*}{$4054\pm3.2$}&$3998.3$ & $1.37$ & $(0,0)$\\
    &$0^{?-}$& &  &$4331.8$ & $6.85$& $(0,0)$\\
    \hline
    $T_{c\bar{c}1}(4200)$&$1^{+-}$&$[cu][\bar{c}\bar{d}]$&$4196^{+35}_{-32}$ &$4318.2$ & $2.91$ &$(1,0)$\\
    \hline
    $T_{c\bar{c}0}(4240)$&$0^{--}$&$[cu][\bar{c}\bar{d}]$&$4239^{+50}_{-21}$ &$4426.7$ & $4.43$ &$(0,0)$\\
    \hline
    $\chi_{c1}(3872)$ & \multirow{4}{*}{$1^{++}$}&\multirow{4}{*}{$[cu][\bar{c}\bar{d}]$}&$3871.64\pm0.06$ & $4129.2$ & $6.67$ &\multirow{4}{*}{$(0,0)$}\\
    $\chi_{c1}(4140)$ & & &$4146.5\pm 0.3$ & $4446.6$ & $7.25$ & \\
    $\chi_{c1}(4274)$ & & &$4286^{+8}_{-9}$ & $4721.1$ & $10.1$&\\
    $\chi_{c1}(4685)$ & & & $4684^{+15}_{-17}$& $4964.2$ & $5.98$&\\
    \hline
    \hline
    \multicolumn{6}{c}{RMS error (17 states, 6 parameters)$=6.26\%.$}
    \end{tabular}
    \caption{Summary of results for cham tetraquarks. Experimental data is read from the PDG site \cite{ParticleDataGroup:2024cfk}. The parameter settings are summarized in Table \ref{tab:five}.}
    \label{tab:four}
\end{table*}    
\end{center} 


As subject testers, we choose the $T_{cc}$ kind of open charm tetraquarks, $T_{c\bar{c}}$-like states, and $\chi_{c1}$ states as hidden charm tetraquarks. In Table \ref{tab:four} we summarize the main results. 


In the case of the open charm sector, we analyzed the $T_{cc}(3875)^+$,  $T^*_{cs0}(2870)$, and $T_{cs0}(2870)^+$ states.  In the former case, we consider two possible parametrizations: supposing $T_{cc}(3875)^+$ as a $S-$wave vector state with mass $3990.2(2.99\%)$ MeV, or as a $S-$wave scalar state with mass $3846(0.74\%)$ MeV. In both cases, we consider $i=1$, accounting for inner dynamical effects. Parameters $\kappa$ and $\alpha$ where chosen using expressions \eqref{kappa_m}  and \eqref{alpha_M}, see appendix \ref{Appx2} for further details. This data is summarized in Table \ref{tab:four}.

\begin{center}
    \begin{table}[h]
        \centering
        \begin{tabular}{c|c}
        \multicolumn{2}{c}{\textbf{Constituent masses Setting in MeV}}\\
        \hline
        \hline
        $m_{u}$ & $339.5$ \\
        $m_d$   & $335.5$ \\
        $m_s$   & $510$ \\
        $m_c$   & $1550$ \\
        $m_l=\frac{m_u+m_d}{2}$ & $337.5$\\
        \hline
        \hline
        \multicolumn{2}{c}{\textbf{Parameter setting for $V_{4Q}(z)$}}\\
        \hline
        \hline
        $\kappa_{c\,\bar{c}}=\kappa\left(m_c\right)$ & $2180.7$ MeV\\
        $\alpha_{c\,\bar{c}}=\alpha\left(m_c\right)$ & $0.5641$\\
        \hline
        $\kappa_{c\,l}=\kappa\left(\frac{m_c+m_l}{2}\right)$ & $1124.7$ MeV\\
        $\alpha_{c\,l}=\alpha\left(\frac{m_c+m_l}{2}\right)$ & $0.2608$ \\
        \hline
        $\kappa_{l\,s}=\kappa\left(\frac{m_l+m_s}{2}\right)$ & $614.5$ MeV\\
        $\alpha_{l\,s}=\alpha\left(\frac{m_l+m_s}{2}\right)$ & $0.034$ \\
        \hline\hline
        \end{tabular}
        \caption{Summary of constituent masses for quarks read from \cite{Scadron:2006dy}, where authors consider for $m_s$ and $m_c$ half of the $\phi(1020)$ and $m_{J/\Psi}$ reported masses. In the second lower part, we wrote the parameter setting for $\kappa$ and $\alpha$ used in the $V_{4Q}(z)$ potential. The first pair, $\kappa_{c\bar{c}}$ and $\alpha_{c\bar{c}}$, correspond for hidden charm states. The last two pairs correspond to the open charm states. These values are calculated with eqns. \eqref{kappa_m} and \eqref{alpha_M}. } 
        \label{tab:five}
    \end{table}
\end{center}

In the case of $T^*_{cs0}(2870)^0$ and $T^*_{cs1}(2900)$, we also fixed the states to be $S-$wave. No contribution coming from the inner structure was considered, so $i=J=0$. In this scenario, $\kappa$ and $\alpha$ were chosen considering a threshold mass given by $\bar{m}=\frac{m_c+m_s}{2}$ and the expressions \eqref{kappa_m} and \eqref{alpha_M}, see apendix \ref{Appx2}. This parameter setting is summarized in Table \ref{tab:five}\\

For the hidden charm case, states $T_{c\bar{c}X1}$, where $X$ accounts for $u$ $c$ or $s$ quarks, can be considered possible tetraquark states if, for neutral states, quantum numbers can be written as $J^{PC}=1^{+-}$, corresponding to a diquark $[cq_1]$ in spin-1 and a antidiquark $[\bar{c}\bar{q}_2]$, with $q_{i}$ being $u$, $d$ or $s$ quarks. These states can also be associated with molecular states. However, experimental evidence is required to support this claiming.\\

Regarding the parameter choice for $\kappa$ and $\alpha$, for hidden charm tetraquarks we will consider a threshold mass given by $\bar{m}=m_c$, see appendix \ref{Appx2} for further details. This choice is supported from phenomenological grounds, as the most dominant decay channel for hidden charm tetraquarks is $ J/\psi$ plus light mesons. Recall that this sort of decay channel is among the most common discovery channels due to clean experimental signatures. However, dominance varies by mass, $J^{PC}$, and inner structure, i.e.,  compact tetraquark vs. molecular picture \cite{Brambilla:2019esw}. Table \ref{tab:four} summarizes the parameter fixing in this case. \\

In the particular case of $T_{c\bar{c}}(4050)^+$, the diquark-antidiquark picture matches better since the $D_0\,\bar{D}^{0*}$ molecule is neutral. Thus, we will theorize two possible configurations: 

\begin{enumerate}
    \item Let us consider that both the diquark and the antidiquark are in a spin-1 configuration. Therefore, it is expected to have $J^{PC}=0^{++}$, implying this state can be modeled as a ground state. 
    \item Another configuration is a spin-0 diquark with an spin-1 antidiquark, allowing $J^{PC}=1^{-+}$ with $L=1$ as possible quantum numbers. 
\end{enumerate}

In both cases, spin effects can be mimicked with the \emph{twist-like} $i$ and $j$ numbers. Results for these fittings are summarized in Table \ref{tab:four}. We found that, as a holographic prediction, $T_{c\bar{c}}(4050)^+$, can be associated with a state having $J^{PC}=1^{-+}$ and mass close to $4215.4(4.06\%)$ MeV.  

A similar situation happens with $T_{c\bar{c}}(4055)^+$, with $J=?^{?-}$ reported by experiments. We can consider two possible configurations $1^{+-}$ and $0^{--}$, following similar spin arguments as we mentioned above. We found that, for this state, the configuration $1^{+-}$ with a mass of $3998.3(1.37\%)$ MeV fits better for $T_{c\bar{c}}(4055)^+$, see Table \ref{tab:four}.  

For $T_{c\bar{c}1}(4200)$, the best fit for the hadron mass is obtained when we include $i=1$ or $j=1$, accounting for one extra Laplacian operator. In the case of the $T_{c\bar{c}0}(4240)$ state, this correction was not included, as the best fit is obtained for $i=j=0$. These results are summarized in Table \ref{tab:four}. 

It is worth mentioning the $\chi_{c1}$ case separately. This set of states can be organized as a single radial Regge trajectory. This hypothesis is consistent with tightly bound states, as these compact structures could allow for excited \emph{metastable} hadrons \cite{Monemzadeh:2014vhh, Tiwari:2022azj}. However, this picture differs from the experimental evidence, suggesting that $\chi_{c1}$ states may have molecular contributions to their structure, given their $J^{PC}=1^{++}$ configuration. This molecular picture consists of a weakly interacting pair of $D^0\bar{D}^{*0}$ mesons held together by a residual \emph{nuclear-like} forces, like pion/kaon exchanges \cite{Wang:2021aql, Chen:2025nkp}. However, for this holographic toy model, we will consider that $\chi_{c1}$ states have an angular distribution, implying that $L=1$. Thus, the proposed tetraquark configuration $[cu][\bar{c}\bar{d}]$ will meet the experimental evidence.   


This hypothesis aligns with interpretations that consider $\chi_{c1}$ states as having a significant molecular component, rather than being purely compact tetraquarks. For the purpose of this holographic toy model, however, we will consider that the $\chi_{c1}$ states are described by the compact tetraquark configuration $[cu][\bar{c}\bar{d}]$ with an angular excitation ($L=1$), which allows the quantum numbers to match the experimental evidence $J^{PC}=1^{++}$.

The formalism presented here, with its foundation in the diquark Bethe-Salpeter equation and the use of local operators with Laplacians, is inherently more suited to modeling compact tetraquark states, which are envisioned as point-like objects in the holographic bulk. While certain states, such as the $\chi_{c1}(3872)$, are also discussed in molecular scenarios, our approach provides a specific, quantitative framework for the compact tetraquark interpretation, yielding predictions that can be tested against future experimental and lattice QCD results.


\section{Conclusions}\label{Concl}
In conclusion, we have developed a robust WKB-based formalism to engineer dilaton fields in holographic QCD models directly from empirical hadron spectra. By solving the inverse problem, we have shown that the observed nonlinearity in the radial Regge trajectories of heavy quarkonia necessitates a departure from the quadratic dilaton of the standard softwall model, leading to a generalized profile $\Phi(z)=\left(\kappa\,z\right)^{2-\alpha}$. The successful application of this method to the charmonium and bottomonium spectra, with RMS errors of 12.2\% and 7.7\% respectively, validates this approach and establishes a direct link between the spectral exponent $\nu$ and the dilaton exponent $\alpha$. This provides a powerful bottom-up recipe where the confining dynamics are not assumed but derived from data.

Furthermore, we have demonstrated the extensibility of this framework by incorporating multiquark states. By augmenting the mesonic confining potential with an additional term ($\tilde{V}(z)$ derived from the diquark Bethe-Salpeter equation, we constructed a composite holographic potential for tetraquarks. The promising results for various open and hidden-charm candidates highlight the potential of this method to describe exotic hadron spectra within a unified holographic picture. This work underscores the value of the WKB inversion technique as a versatile tool for bottom-up model building, extending beyond mesons and baryons to map the increasingly complex landscape of exotic hadrons onto a dual gravitational theory, providing a quantitative bridge between QCD phenomenology and gravitational duals.

\begin{acknowledgments}
M. A. Martin Contreras wants to acknowledge the financial support provided by the National Natural Science Foundation of China (NSFC) under grant No. 12350410371. A. V is partially supported by Centro de Física Teórica de Valparaíso (CEFITEV) and by FONDECYT (Chile) under grant No. 1251106.
\end{acknowledgments}

\appendix
\section{Determination of $\sigma_c$ from the Cornell Potential}\label{ap1} 
Let us start from the time-independent Schrödinger equation for the Cornell potential \cite{Quigg:1979vr, Lucha:1991vn} used to solve heavy quarkonium spectroscopy:

\begin{equation}
    \left[-\frac{1}{2\,\mu}\nabla^2+U_c(r)\right]\Psi\left(\vec{r}\right)=E\,\Psi\left(\vec{r}\right)
\end{equation}

\noindent with $\mu$ defined as the \emph{quark reduced mass}, and the Cornell potential $U_c(r)$ is written as 

\begin{equation}\label{cornell}
    U_c(r)=-\frac{4}{3}\frac{\alpha_s}{r}+\sigma\,r+V_0
\end{equation}

\noindent with $\alpha_s$ the strong coupling constant, $\sigma$ the string tension (in units of GeV$^2$), and $V_0$ is a regularization constant. Let us consider the linear confinement term only, focusing on the WKB analysis. 

In this scenario, the eigenvalues of the Schrödinger equation are the binding energies $E_n$, related to the hadron mass $M_n$ as 

\begin{equation}
    M_n=m_{q_1}+m_{q_2}+E_n-V_0
\end{equation}

The eigenvalues $E_n$ for the Cornell potential can be calculated in terms of zeroes $\epsilon_n$ of Airy polynomials $Ai(x)$, as

\begin{equation}
    E_n=-\left(\frac{\sigma^2}{2\,\mu}\right)^{1/3}\,\epsilon_n.
\end{equation}

We can use the expression above and compute the mass shift between the ground and first excited states:

\begin{equation}\label{mass-shift}
    M_2-M_1=E_2-E_1=\left(\frac{\sigma^2}{2\,\mu}\right)^{1/3}\left(\epsilon_2-\epsilon_1\right).
\end{equation}

For the charmonium case, we consider $J/\Psi$ and $\Psi'$ to evaluate the mass shift. Therefore, we can extract the value of $\sigma_c$, yielding:

\begin{equation}
    \sigma_c=\sqrt{2\,\mu_c}\left(\frac{M_{\Psi'}-M_{J/\Psi}}{\epsilon_2-\epsilon_1}\right)^{3/2}=0.243\,\text{GeV}^2,
\end{equation}

\noindent where $\mu_c=m_c$ for these $c\bar{c}$ vector states. Notice that, when the one-gluon exchange term ($\sim -1/r$), i.e., including low $n$ effects, we obtain the standard value for $\sigma=0.18$ GeV$^2$ for $\alpha\sim0.35$ for fitting heavy quarkonia in NRQCD \cite{Lucha:1991vn}. However, we chose the former value, $\sigma_c=0.243$ GeV$^2$, since we consider highly excited radial states.

\section{Non-quadratic dilaton for heavy quarkonium}\label{Appx2}
The non-quadratic dilaton $\Phi(z)=\left(\kappa\,z\right)^{2-\alpha}$ formulation can be extended to describe the \emph{isoscalar meson family spectrum}, composed by $\omega$, $\phi$, $\psi$ and $\Upsilon$ mesons, with quantum numbers given by $I^G(J^{PC})=0^+(1^{--})$, see further details in \cite{MartinContreras:2020cyg,  MartinContreras:2021bis, MartinContreras:2023oqs}. The main hypothesis behind this dilaton profile is the expected effect of the constituent mass deviating from the quadratic behavior of the softwall model dilaton \cite{Karch:2006pv}, which is intended for light, unflavored mesons. Therefore, for the isoscalar family, we expect that a general radial Regge trajectory, given by

\begin{equation}
    M_n^2=a(n+b)^\nu,
\end{equation}

\noindent where $a$ is the slope (usually flavor dependent) in energy units, $b$ is the intercept, and $\nu$ accounts for the linearity deviation. 

This idea is supported by the Bethe-Salpeter analysis \cite{Chen:2018bbr} when including the quark mass in the radial trajectory:

\begin{equation}
    \left(M_n^2-m_1-m_2\right)^2=a(n+b).
\end{equation}

Following the Bethe-Saltpeter approximation, variations in linearity are expected to be induced by the constituent mass. Within the context of heavy quarkonium studies, we can derive $M_n^2\propto n^{2/3}$, indicating that intermediate constituent masses might correspond to values of $\nu$ that span between one and $2/3$. However, in the bottom-up approach, the upper bound reached in the heavy quark mass is smaller than 2/3. 

For bottomonium states, unlike the Bethe-Salpeter theory, it is anticipated that the exponent $\nu$ should be less than $2/3$. This particular phenomenon implies that the constituent quark mass likely modulates $\nu$. At the holographic level, this approach will enable us to derive insights into the internal configuration of the relevant hadron.

When considering the isoscalar family, it is possible to define a relation between $\kappa$ and $\alpha$ with the constituent mass $\bar{m}$ univocally, since for each of the four spectra it is possible to find a pair $(\kappa,\,\alpha)$, which encodes the nonlinearity effect, that emerges from the quark constituent mass \cite{MartinContreras:2020cyg} . Such relations are found to be written as \cite{MartinContreras:2024gsk}:

\begin{eqnarray} \label{kappa_m}
    \kappa (\bar{m}) &=& 30.613-30.129\,e^{-0.0241\, \bar{m}^{2}},\\ \label{alpha_M}
    \alpha(\bar{m}) &=& 0.8965-0.9315\,e^{-0.4233\, \bar{m}^{2}},\\
    \mathcal{R}^2&=&0.999\,\,\text{for both fits.}
\end{eqnarray}

These relations show how the slope $\kappa$ and $\alpha$ increase with the constituent mass \emph{threshold}. When the heavy quark mass limit is taken,  notice that $\kappa_{HQ}\to 30.61$ GeV, higher than the slope for the bottomium, hadrons with the heaviest constituent mass,  and $\alpha_{HQ}\to 0.1876$, smaller than fitted value for the bottomonium spectrum, $0.193$. 

The set of equations \eqref{kappa_m} and \eqref{alpha_M} can be used as \emph{calibration curves}, allowing us to compute $(\kappa,\alpha)$ provided a parametrization for the constituent mass $\bar{m}$. Since we want to address different hadronic states beyond the conventional ones, we employ the following \emph{stoichiometric} relations:

\begin{eqnarray}
    \bar{m}&=&\sum_{i=1}^N(P^{Q_i}_i\,\bar{m}_{Q_i}+P^{G_i}_i\,m_{G_i}+P^{M_i}_i\,m_{M_i}),\\
    1&=&\sum_{i=1}^N\,(P^{Q_i}_i+P^{G_i}_i+P^{M_i}_i). 
\end{eqnarray}

\noindent where $P^{Q_i}_i$, $P^{G_i}_i$ and $P^{M_i}_i$ are \emph{constituent weights} that fixes the contributions to the \emph{average constituent mass} from quarks, gluons and other meson cores respectively. 

From these relations, it is possible to define univocally $\kappa$ and $\alpha$ for hadronic states. In the context of standard mesons, with no other contributions coming from gluons or inner mesonic cores, we choose the \emph{average constituent mass} as 

\begin{equation}
    \bar{m}_\text{Meson}=\frac{m_{Q_1}+m_{Q_2}}{2}.
\end{equation}

\noindent where $P^{Q_i}_i=1/2$ in this particular case. This mass is used as input into the stoichiometric relations \eqref{kappa_m}
and \eqref{alpha_M} to compute the corresponding $\kappa$ and $\alpha$. 

\section{Regge Trajectories from Quadratic Spinless Bethe-Salpeter equation}\label{app2}

In the context of the Quadratic Spinless Bethe-Salpeter equation, Refs. \cite{Feng:2023txx, Chen:2023cws} have proposed a general, i.e., nonlinear, form for the Regge trajectories written as

\begin{equation}\label{gen-regge}
    M=\beta_x\,\left(x+c_{0x}\right)^\nu+m_R,
\end{equation}

\noindent where $x$ represents the orbital angular momentum $L$ or the radial quantum number $n$. The quantities $\beta_x$ and $\nu$ are parameters determined from the QSSE. 

For the case of diquarks, these states emerge as relativistic bound states calculated from the  QSSE master equation:

\begin{equation}
    M^2\,\Psi(\textbf{r})=M_0^2\,|\Psi(\textbf{r})+\mathcal{U}\,\Psi(\textbf{r}), 
\end{equation}

\noindent where we have defined the following expressions 

\begin{eqnarray}
    M_0&=&\omega_1+\omega_2\\
    \mathcal{U}&=&M_0\,V_{QQ'}+V_{QQ'}\,M_0+V_{QQ'}^2\\
    \omega_i&=&\sqrt{m_i^2+\textbf{p}^2}
\end{eqnarray}

\noindent corresponding with the sum of relativistic kinetic energies $\omega_i$, the quark-quark interaction term in the diquark cluster, and $\omega_i$ accounts for the quark energy. Finally,  $\Psi(\textbf{r})$ is the relativistic wave function. See \cite{Baldicchi:2007ic} for further details.

For the  $V_{QQ'}$, i.e, the potential inside the diquark cluster,  it is fixed to be half of the Cornell potential \eqref{cornell}. The quark-quark interaction potential accounts for the color $\bar{3}_c$ representation, which is attractive. 

To make numerical analysis, we choose the non-relativistic limit $m_1,\,m_2\gg \left|\textbf{p}\right|$, we obtain  

\begin{equation}
    M_0\approx m_1+m_2+\left(\frac{1}{2\,m_1}+\frac{1}{2\,m_2}\right)\,\textbf{p}^2=m_1+m_2+\frac{\textbf{p}^2}{2\,\mu},
\end{equation}

\noindent with $\mu$ defined as the reduced mass. 

Thus, the QSSE, when keeping leading-order terms, reduces to a Schrodinger-like limit written as

\begin{multline}
    M^2\,\Psi(\textbf{r})\approx\left[\left(m_1+m_2\right)^2+\frac{m_1+m_2}{2\,\mu}\,\textbf{p}^2\right.\\
\left.+2\,\left(m_1+m_2\right)\,V_{QQ'}\right]\,\Psi(\textbf{r}).
\end{multline}

Then, we apply the WKB approach for bound states, the so-called \emph{Bohr-Sommerfeld quantization}, to the potential $V_{QQ'}$ \cite{Quigg:1979vr}: 

\begin{equation}
    \oint{p_r\,dr}=2\,\pi\,\left(n+\frac{1}{2}\right),
\end{equation}

\noindent where $p_r$ is the radial momentum. Notice that near the potential turning points, the dominant part of the potential is $\sigma \,r$. Thus, we evaluate the integral as:

\begin{equation}
    E_n\sim\left(\frac{\sigma^2}{4\,\mu}\right)^{1/3}\,n^{2/3}.
\end{equation}

Similarly, for large orbital angular momentum, we consider the centrifugal term, leading to:

\begin{equation}
    E_L\sim\left(\frac{\sigma^2}{4\,\mu}\right)^{1/3}\,L^{2/3}.
\end{equation}

Finally, collecting all of these results, we can write the Regge trajectory as

\begin{eqnarray}\notag
M^2&\sim&\left(m_1+m_2\right)^2+2\,\left(m_1+m_2\right)\left[\text{kinetic+potential}\right]  \\
&\sim&\left(m_1+m_2\right)^2+2\,\left(m_1+m_2\right)\,\left(\frac{\sigma^2}{4\,\mu}\right)^{1/3}x^{2/3},
\end{eqnarray}

\noindent motivating the generalized Regge ansatz proposed in \eqref{gen-regge}, when we identify

\begin{eqnarray}
    m_R&=&m_1+m_2+\varepsilon_Q\\
    \beta_x&=&c_{f_x}\,c_x\,c_Q\\
    c_Q&=&\left(\frac{\sigma^2}{4\,\mu}\right)^{1/3},
\end{eqnarray}

\noindent where we have included $\varepsilon_Q$ defined from the Cornell-Potential regulator $V_0$, the parameters $c_{f_x}$ and $c_x,$ fixed by numerical fitting, and $c_Q$ encloses the confinement and reduced mass effects. Following \cite{Feng:2023txx}, for open-charm diquarks in $S$-wave, $[cc]^{\textbf{3}_c}_{1^3s_1}$, with a constituent mass of $m_c=1.55$ GeV  and $\sigma_c=0.18$ GeV$^2$, brings an output of $c_{0l}=0.337$, $c_{fl}=1.17$ and $c_l=3/2$, for an diquark mass of $m_{[cc]}\approx3.14$ GeV. 

It is worth noting that this fit applies to large $n$ and $L$. However, the Coulombian part has to be included to match the low $n$, $L$. Similar arguments follow for the holographic counterpart. 
\bibliography{apssamp}

\end{document}